\documentclass[nofootinbib]{revtex4}

\usepackage{amsfonts}

\begin{document}

\title{Quantum Gravity phenomenology and metric formalism}

\author{Niccol\'{o} Loret}

\address{Institut Rudjer Boskovich,\\
 10000 Zagreb, Bijenichka cesta 54, Croatia\\
E-mail: Niccolo.Loret@irb.hr}

\author{Leonardo Barcaroli} 

\address{Dipartimento di Fisica, Universit\'{a} di Roma ``La Sapienza",\\
 P.le A. Moro 2, 00185 Roma, Italy\\
E-mail: leonardo.barcaroli@roma1.infn.it}

\author{Giulia Gubitosi} 

\address{Theoretical Physics Blackett Laboratory, Imperial College,\\
 London, SW7 2AZ, United Kingdom\\
E-mail: g.gubitosi@imperial.ac.uk}

\begin{abstract}
In this proceedings for the MG14 conference, we discuss the construction of a phenomenology of Planck-scale effects in curved spacetimes, underline a few open issues and describe some perspectives for the future of this research line.
\end{abstract}


\maketitle

\section{Introduction}

In the last decade Quantum Gravity phenomenology developed quickly, investigating many different possible manifestations of tiny  Planck-scale effects at macroscopical level. One of the focal points of this research line is a class of modifications of the dispersion relation for free-moving particles \cite{grbgac,gacdsr1a,gacdsr1b}, parametrized through two dimensionsless  parameters $\eta$ and $\alpha$, in terms of the Planck scale $M_P$ (in this paper we use units such that $c=1$):
\begin{equation}
p_0^2-m^2=p^2\left(1+\eta\left(\frac{p_0}{M_P}\right)^\alpha\right)\,.\label{genericMDR}
\end{equation}
Since the form of the dispersion relation is an important landmark in special relativity, its modification suggests scenarios of Lorentz symmetries violation or deformation at energies close to $M_P$. However the presence of amplifying mechanisms, such as thresholds effects \cite{aloisio} or cosmological distances of propagation \cite{grbgac,gacdsr1a,gacdsr1b,JacPir,invacuo}, as well as high-precision experiments \cite{gacdsr1b,jacLibMatt,FlaCarCor,weakness}, makes it possible to consider a Quantum Gravity phenomenology at energies much lower than $M_P$. When the effects of Planck-scale deformations of particles' dispersion relation are considered in an astrophysical or cosmological context, it is necessary to correctly introduce spacetime curvature. This is needed in order to take into account in the right way the gravitational redshift contribution on the propagating particle observables, or on the interactions we intend to study. It would be, then, an important contribution to this research-line the construction of a metric framework formalising the interplay between Planck-scale effects and spacetime curvature. Some issues related to this task are discussed in these proceedings. 

\section{Rainbow Metrics}

One of the most studied features of QG phenomenology, which can be obtained in Planck-scale deformed relativistic models, formalized through a modified dispersion relation (MDR) like the one in eq. (\ref{genericMDR}), is the momentum dependence of massless particles' velocity. This feature suggests that the light cone structure should be modified accordingly with the particles' velocity expression. The first attempt to obtain a coherent metric formalism describing this kind of physical effect as a momentum dependence of the metric itself was made in \cite{rainbow} by Smolin and Magueijo. In this approach free field theories have plane wave solutions, even though the 4-momentum they carry satisfies deformed dispersion relations which one can express as
\begin{equation}
p_0^2 f^2\left(\frac{p_0}{M_P}\right)-(p\cdotp p) g^2\left(\frac{p_0}{M_P}\right)=m^2,\label{RainbowSmolin} 
\end{equation} 
where $f(p)$ and $g(p)$ are two generic functions depending on momenta. Then spacetime metric should be modified according to the energy of the particle we use to probe it. The MDR described in (\ref{RainbowSmolin}) can be realized by the action of a nonlinear map from momentum space to itself, denoted as $U\,:\,\mathcal{P}\rightarrow\mathcal{P}$, given by
$$
U\cdotp(p_0,p_i)=(U_0,U_i)=\left(f\left(\frac{p_0}{M_P}\right)p_0,g\left(\frac{p_0}{M_P}\right) p_i\right),
$$
which would imply that momentum space has a nonlinear norm, given by
\begin{equation}
|p|^2=\eta^{ab}U_a(p)U_b(p).
\end{equation}
If one still wants to have plane wave solutions for free fields, since momentum transforms nonlinearly, the contraction between position and momentum, which can be also formalised as $\{p_\mu , x^\nu\}=\delta_\mu^\nu$, must remain linear. Smolin and Magueijo suggested that in case momentum transforms nonlinearly, this can be obtained asking that
\begin{equation}
\zeta^{\alpha\gamma}\tilde{g}_{\gamma\beta}=\delta_\beta^\alpha,
\end{equation}
where $\zeta^{\alpha\gamma}$ is the metric of momentum space and $\tilde{g}_{\gamma\beta}$ is the so called "Rainbow metric", such that the spacetime interval
\begin{equation}
ds^2=\tilde{g}_{\gamma\beta}(p)dx^\gamma dx^\beta=\frac{(dx^0)^2}{f^2(p)}-\frac{(dx^i)^2}{g^2(p)}\label{noninvlinel}
\end{equation}
is explicitely energy-dependent.\\
This construction, however brilliant from the phenomenological point of view (for it has paved the road to the study of a broad number of new Planck-scale effects), suffers from the lack of invariance of the "invariant" line-element (\ref{noninvlinel}). This issue can be explained for example choosing a simple parametrization for the MDR, in which $f^2(p)=1$ and $g^2(p)=1+\ell p_0$\footnote{This could be also in general obtained by fixing the parameters in (\ref{genericMDR}) as $\alpha=1$ and $\eta/M_P=\ell$}, and the rainbow spacetime metric can be represented\footnote{In this framework we usually work at first order in the parameter $\ell$.} in 1+1D as
\begin{equation}
\tilde{g}_{\alpha\beta}=\left(\begin{array}{cc}
1 & 0\\
0 & -(1-\ell p_0)
\end{array}\right)\, .
\end{equation}
However as we stated before, the line-element that we obtain from this metric formalism
\begin{equation}
ds^2=\tilde{g}_{\alpha\beta}(p)dx^\alpha dx^\beta\, ,
\end{equation}
is not invariant under the same symmetries that let the MDR unchanged under boost transformation. In fact the representation of the boost generator of this simple model being  
\begin{equation}
\mathcal{N}=x^0 p_1 + x^1\left(p_0-\ell p_0^2-\frac{\ell}{2}p^2\right)\, ,\label{boostrapp}
\end{equation}
and letting the boost act on a generic observable $\mathcal{A}$ through Poisson brackets
\begin{equation}
\mathcal{N}\rhd \mathcal{A}=\mathcal{A}'=\mathcal{A}-\xi\{\mathcal{N},\mathcal{A}\}\,,
\end{equation}
where $\xi$ is the rapidity parameter, we can express the line-element transformation as
\begin{equation}
(ds')^2=(d(x^0)')^2-(1-\ell (p_0)')(d(x^1)')^2=\tilde{g}_{\mu\nu}dx^\mu dx^\nu+\ell \xi ((dx^1)^2 p_1-2 dx^0 dx^1 p_0) \neq ds^2\, .\nonumber\label{invarianteuncorno}
\end{equation}
This last equation shows how Rainbow metrics cannot formalise a Planck-scale deformed symmetries framework, and therefore that they are mostly suited to describe Lorentz symmetries breakdown scenarios.

\section{A Further Formalization: Relative Locality}

A new attempt to obtain a coherent framework for relativistic Planck-scale-deformed dispersion relations has been made in the so called Relative Locality approach \cite{bob,principle,kbob,transverse,lateshift}. In Relative Locality the dispersion relation is obtained from the momentum-space invariant line-element:
\begin{equation}
{\cal C}=\int_0^1 \zeta^{\mu\nu}(P)\dot{P}_\mu\dot{P}_\nu\;d\tau\, ,\label{calcmom}
\end{equation}
in which $\tau$ is the variable with which we parametrise the geodesic $P(\tau)$ connecting the point $p$ to the origin of momentum-space. Therefore a deformation of the dispersion relation can be seen as the effect of curvature of momentum-space. For example, from a deSitter-like curvature expressed at first order in $\ell$ by the metric
\begin{equation}
\zeta^{\alpha\beta}=\left(\begin{array}{cc}
1 & 0\\
0 & -(1+2\ell p_0)
\end{array}\right)\, ,\label{mommetr}
\end{equation}
can be obtained\cite{trevisan} the Casimir operator of the spacetime symmetry algebra
\begin{equation}
\mathcal{C}(p)=p_0^2-p_1^2-\ell p_0 p_1^2\,,
\end{equation}
whose on shell-relation $\mathcal{C}=m^2$ gives a dispersion relation of the same kind of eq. (\ref{RainbowSmolin}) with the already mentioned parametrisation $f^2(p)=1$ and $g^2(p)=1+\ell p_0$. The spacetime line element we obtain imposing spacetime metric to be the inverse of momentum-space one (\ref{mommetr})
\begin{equation}
ds^2=\zeta_{\lambda\mu}(p)dx^\lambda dx^\mu\,,
\end{equation}
is invariant under the action of boost (\ref{boostrapp}). Thus when considering deformations of the flat spacetime dispersion relation, it is possible to find a well defined Planck-scale deformed metric of spacetime \cite{SpecRelLoc}. However turning on spacetime curvature gives rise to a few complications. We can show this issue through the generalisation (see {\it exempli gratia} \cite{GiuqdS,ChridS}) of metric (\ref{mommetr}) when spacetime has deSitter geometry:
\begin{equation}
\zeta^{\alpha\beta}_{H\ell}=\left(\begin{array}{cc}
1 & 0\\
0 & -e^{-2 H x^0}(1+2\ell p_0)\\
\end{array}\right)\,,\label{metr1+1D}
\end{equation} 
in which we assume $H\sim\sqrt{\Lambda/3}$ being $\Lambda$ the cosmological constant. This metric allows us to find both the Casimir operator and the particles' worldlines. The Casimir operator can be found using formula (\ref{calcmom}) to be: 
\begin{equation}
{\cal C}_{H\ell}(p,x)=p_0^2-p_1^2 e^{-2 H x^0}-\ell p_0 p_1^2 e^{-2 H x^0}\,.\label{ClH}
\end{equation}
Worldlines are also easy to find using this formalism \cite{unicorns}, observing that:
\begin{equation}
x^1(x^0)-\bar{x}^1=\int_{\bar{x}^0}^{x^0} \sqrt{\frac{\zeta_{00}}{-\zeta_{11}}} dx^0=\frac{e^{-H x^0}-e^{-H \bar{x}^0}}{H}+\ell\bar{p}_0e^{-H x^0}\frac{\sinh(H(x^0-\bar{x}^0))}{H}\,,\label{qworld1}
\end{equation}
where $\bar{p}_0$, $\bar{x}^0$ and $\bar{x}^1$ are the initial conditions for the particle motion. 
Even if in this formalism the Casimir and worldlines are invariant under some generalised boost transformations, it is not clear whether this is the case also for the momentum space and the spacetime line element $ds^2_{H\ell}$, which are derived from the metric (\ref{metr1+1D}). In fact, since the metric is now explicitly $(x,\dot{x})$-dependent, it may be not possible anymore to obtain a formalization of the invariant line-elements in terms of the coordinates differentials. This may not be a problem, since line-element and metric are not something we observe, but just useful tools we infer from particles' dynamics. At the same time metric formalism is fundamental in many aspects of General Relativity and we would like to preserve it also in it's deformed version that we may want to formulate one day.  

\section{Conclusions}
The situation we described in this proceedings shows some results achieved and many open issues to solve for what concerns the use of metrics to formalise spacetime-symmetries deformation models. The road ahead presents many choices. We could focus in understanding under which assumptions can we inject all this model deformation just in one geometric object, leaving unchanged the remaining relativistic structure of he theory. In this sense tetradic formalism has proved to be very fruitful \cite{KowaJack2}, but at the same time how can we build a relativistic framework on the invariant line-element if at the end this one would not result a reliable landmark? A drastic solution to this problem could be to renounce to rely on metric formalism and investigate on Finsler geometry \cite{Finsler} as a way to formalise Planck-scale deformed relativistic symmetries, as an alternative to the implementation of  those deformations directly on Riemannian geometry, since we don't know yet if such an approach would or would not work. Another approach could be to build from scrath a theory which takes into account a complete scenario of curved phase-space \cite{HamGeo}, building a metric formalism from phase space symmetries.


\begin{thebibliography}{0}

\bibitem{grbgac}
G. Amelino-Camelia \emph{et~alii},
 Nature \textbf{393}, 763-765 (1998).
 
 \bibitem{gacdsr1a}
G. Amelino-Camelia,
 Int. J. Mod. Phys. D {\bf 11}, 35-60 (2002).

\bibitem{gacdsr1b}
G. Amelino-Camelia,
 Phys. Lett. B {\bf 510}, 255-263 (2001).

\bibitem{jacLibMatt}
T. Jacobson, S. Liberati, D. Mattingly,
 Annals Phys. {\bf 321}, 150-196 (2006).
 
\bibitem{aloisio}
R. Aloisio, P. Blasi, P. L. Ghia and A. F. Grillo,
 Phys. Rev. D {\bf 62}, 053010 (2000).
 
\bibitem{JacPir}
U. Jacob and T. Piran, 
 Nature Phys. {\bf 3}, 87 (2007).
 
\bibitem{invacuo}
G.~Amelino-Camelia, G.~D'Amico, G.~Rosati and N.~Loret,
  Nat.\ Astron.\  {\bf 1}, 0139 (2017).

\bibitem{FlaCarCor}
F. Mercati, D. Maz\'{o}n, G. Amelino-Camelia, J. M. Carmona, J. L. Cort\'{e}s, J. Indur\'{a}in, C. Laemmerzahl, G. M. Tino,
 Class. Quant. Grav. {\bf 27}, 215003 (2010). 
 
\bibitem{weakness}
G.~Amelino-Camelia, G.~Gubitosi, N.~Loret, F.~Mercati and G.~Rosati,
  EPL {\bf 99}, no. 2, 21001 (2012).
 
\bibitem{rainbow}
J. Magueijo, L. Smolin,
 Class. Quant. Grav. {\bf 21}, 1725-1736 (2004).

\bibitem{bob}
G.~Amelino-Camelia, M.~Matassa, F.~Mercati and G.~Rosati,
 Phys. Rev. Lett. {\bf 106}, 071301 (2011).

\bibitem{principle}
G.~Amelino-Camelia, L.~Freidel, J.~Kowalski-Glikman, L.~Smolin,
 Phys.~Rev.~D {\bf 84}, 084010 (2011).

\bibitem{kbob}
 G.~Amelino-Camelia, N.~Loret and G.~Rosati,
 Phys. Lett. B {\bf 700}, 150 (2011).

\bibitem{transverse}
G. Amelino-Camelia, L. Barcaroli and N. Loret,
 Int. J. of Theoret. Phys. {\bf 51}, 3359 (2012).

\bibitem{lateshift} 
G. Amelino-Camelia, L. Barcaroli, G. Gubitosi, N. Loret,
 Class. Quant. Grav. {\bf 30}, 235002 (2013).

\bibitem{trevisan}
G.~Amelino-Camelia, M.~Arzano, J.~Kowalski-Glikman, G.~Rosati and G.~Trevisan,
 Class. Quantum Grav. {\bf 29}, 075007 (2012).

\bibitem{SpecRelLoc}
N. Loret,
 Phys. Rev. D {\bf 90}, 124013 (2014).
 
\bibitem{GiuqdS}
A. Marcian\'{o}, G. Amelino-camelia, N. Rossano-Bruno, G. Gubitosi, G. Mandanici, A. Melchiorri,
 JCAP {\bf 1006}, 030 (2010). 

\bibitem{ChridS}
L.~Barcaroli, L.~K.~Brunkhorst, G.~Gubitosi, N.~Loret and C.~Pfeifer,
  Phys.\ Rev.\ D {\bf 95}, no. 2, 024036 (2017).

\bibitem{unicorns}
I.~P.~Lobo, N.~Loret and F.~Nettel,
  Eur.\ Phys.\ J.\ C {\bf 77}, no. 7, 451 (2017).

\bibitem{KowaJack2}
F. Cianfrani, J. Kowalski-Glikman, G. Rosati,
 Phys. Rev. D {\bf 89}, 044039 (2014).

\bibitem{Finsler}
G. Amelino-Camelia, L. Barcaroli, G. Gubitosi, S. Liberati, N. Loret,
 Phys. Rev. D {\bf 90}, 125030 (2014).
 
\bibitem{HamGeo}
L. Barcaroli, L. K. Brunkhorst, G. Gubitosi, N. Loret, C. Pfeifer,
 Phys. Rev. D {\bf 92}, 084053 (2015).

\end{thebibliography}
\end{document}